\documentclass[a4paper, 11pt]{article}
\pdfoutput=1
\usepackage{jheppub}

\newcommand{\be}{\ensuremath{\beta} }
\newcommand{\ga}{\ensuremath{\gamma} }
\newcommand{\de}{\ensuremath{\delta} }
\newcommand{\cC}{\ensuremath{\mathcal C} }
\newcommand{\cN}{\ensuremath{\mathcal N} }
\newcommand{\cO}{\ensuremath{\mathcal O} }
\newcommand{\gc}{\ensuremath{g_c^2} }
\newcommand{\gGF}{\ensuremath{g_{\rm GF}^2} }

\newcommand{\gstar}{\ensuremath{g_{\star}^2} }
\newcommand{\gtc}{\ensuremath{\widetilde g_c^2} }
\newcommand{\gtGF}{\ensuremath{\widetilde g_{\rm GF}^2} }
\newcommand{\gtopt}{\ensuremath{\widetilde g_{\rm opt}^2} }
\newcommand{\topt}{\ensuremath{\tau_{\rm opt}} }
\newcommand{\MSbar}{\ensuremath{\overline{\mbox{MS}}} }
\newcommand{\vev}[1]{\ensuremath{\left\langle #1 \right\rangle} }
\newcommand{\refcite}[1]{ref.~\cite{#1}}
\newcommand{\eq}[1]{eq.~\ref{#1}}
\newcommand{\fig}[1]{figure~\ref{#1}}

\title{Improving the continuum limit \\ of gradient flow step scaling}

\author{Anqi~Cheng,$^1$}
\author{Anna~Hasenfratz,$^1$}
\author{Yuzhi~Liu,$^1$}
\author{Gregory Petropoulos$^1$}
\author{and David~Schaich$^{1, 2}$}
\affiliation{$^1$Department of Physics, University of Colorado, Boulder, CO 80309, USA}
\affiliation{$^2$Department of Physics, Syracuse University, Syracuse, NY 13244, USA}
\emailAdd{anna@eotvos.colorado.edu}

\abstract{ 
  We introduce a non-perturbative improvement for the renormalization group step scaling function based on the gradient flow running coupling, which may be applied to any lattice gauge theory of interest.  Considering first SU(3) gauge theory with $N_f = 4$ massless staggered fermions, we demonstrate that this improvement can remove $\cO(a^2)$ lattice artifacts, and thereby increases our control over the continuum extrapolation.  Turning to the 12-flavor system, we observe an infrared fixed point in the infinite-volume continuum limit.  Applying our proposed improvement reinforces this conclusion by removing all observable $\cO(a^2)$ effects.  For the finite-volume gradient flow renormalization scheme defined by $c = \sqrt{8t} / L = 0.2$, we find the continuum conformal fixed point to be located at $\gstar = 6.2(2)$.
}

\begin{document}
\maketitle
\flushbottom

\section{Introduction} 
Asymptotically-free SU($N$) gauge theories coupled to $N_f$ massless fundamental fermions are conformal in the infrared if $N_f$ is sufficiently large, $N_f \geq N_f^{(c)}$.
Their renormalization group (RG) \be functions possess a non-trivial infrared fixed point (IRFP) where the gauge coupling is an irrelevant operator.
Although this IRFP can be studied perturbatively for large $N_f$ near the value at which asymptotic freedom is lost~\cite{Caswell:1974gg, Banks:1981nn}, as $N_f$ decreases the fixed point becomes strongly coupled.
Systems around $N_f \approx N_f^{(c)}$ are particularly interesting strongly-coupled quantum field theories, with non-perturbative conformal or near-conformal dynamics.
Their most exciting phenomenological application is the possibility of a light composite Higgs boson from dynamical electroweak symmetry breaking~\cite{Fodor:2012ty, Matsuzaki:2012xx, Appelquist:2013sia, Fodor:2014pqa, Aoki:2014oha}.
Due to the strongly-coupled nature of these systems, lattice gauge theory calculations are a crucial non-perturbative tool with which to investigate them from first principles.
Many lattice studies of potentially IR-conformal theories have been carried out in recent years (cf.~the recent reviews~\cite{Neil:2012cb, Giedt:2012it} and references therein).
While direct analysis of the RG \be function may appear an obvious way to determine whether or not a given system flows to a conformal fixed point in the infrared, in practice this is a difficult question to address with lattice techniques.
In particular, extrapolation to the infinite-volume continuum limit is an essential part of such calculations.

In the case of SU(3) gauge theory with $N_f = 12$ fundamental fermions, several lattice groups have investigated the step scaling function, the discretized form of the \be function.
To date, these studies either did not reach a definite conclusion~\cite{Hasenfratz:2010fi, Lin:2012iw} or may be criticized for not properly taking the infinite-volume continuum limit~\cite{Appelquist:2007hu, Appelquist:2009ty, Hasenfratz:2010fi, Hasenfratz:2011xn, Itou:2012qn, Petropoulos:2013gaa}.
At the same time, complementary numerical investigations have been carried out, considering for example the spectrum, or bulk and finite-temperature phase transitions~\cite{Deuzeman:2009mh, Fodor:2011tu, Appelquist:2011dp, DeGrand:2011cu, Cheng:2011ic, Cheng:2013eu, Fodor:2012uw, Fodor:2012et, Aoki:2012eq, Aoki:2013zsa, Jin:2012dw}.
The different groups performing these studies have not yet reached consensus regarding the infrared behavior of the 12-flavor system.

Our own $N_f = 12$ results favor the existence of a conformal IRFP, which we observe in Monte Carlo RG studies~\cite{Hasenfratz:2011xn, Petropoulos:2013gaa}.
Our zero- and finite-temperature studies of the lattice phase diagram show a bulk transition consistent with conformal dynamics~\cite{Schaich:2012fr, Hasenfratz:2013uha}.
From the Dirac eigenvalue spectrum~\cite{Cheng:2013eu}, and from finite-size scaling of mesonic observables~\cite{Cheng:2013xha}, we obtain consistent predictions for a relatively small fermion mass anomalous dimension: $\ga_m^{\star} = 0.32(3)$ and 0.235(15), respectively.
While this conclusion, if correct, would render the 12-flavor system unsuitable for composite Higgs phenomenology, we consider $N_f = 12$ to remain an important case to study.
Considerable time and effort has already been invested to obtain high-quality lattice data for the 12-flavor system.
Until different methods of analyzing and interpreting these data can be reconciled -- or the causes of any remaining disagreements can be clarified -- it will not be clear which approaches are most reliable and most efficient to use in other contexts.

The recent development of new running coupling schemes based on the gradient flow~\cite{Luscher:2009eq, Luscher:2010iy, Fodor:2012td, Fodor:2012qh, Fritzsch:2013je} provides a promising opportunity to make progress.
In this work we investigate step scaling using the gradient flow running coupling.\footnote{We are aware of two other ongoing investigations of the $N_f = 12$ gradient flow step scaling function, by the authors of \refcite{Fodor:2012td} and \refcite{Lin:2012iw}.}
We begin by introducing a non-perturbative improvement to this technique, which increases our control over the continuum extrapolation by reducing the leading-order cut-off effects.
While this improvement is phenomenological in the sense that we have not derived it systematically through a full improvement program, it is generally applicable to any lattice gauge theory of interest and can remove all $\cO(a^2)$ cut-off effects.
We illustrate it first for 4-flavor SU(3) gauge theory, a system where the running coupling has previously been studied with both Wilson~\cite{Tekin:2010mm} and staggered~\cite{PerezRubio:2010ke, Fodor:2012td, Fodor:2012qh} fermions.
We then turn to $N_f = 12$, where we show that the infinite-volume continuum limit is well defined and predicts an IRFP.
In both the 4- and 12-flavor systems, our improvement can remove all observable $\cO(a^2)$ effects, despite the dramatically different IR dynamics.
We conclude with some comments on other systems where improved gradient flow step scaling may profitably be applied.

\section{Improving gradient flow step scaling} 
The gradient flow is a continuous invertible smearing transformation that systematically removes short-distance lattice cut-off effects~\cite{Luscher:2009eq, Luscher:2010iy}.
At flow time $t = a^2 t_{\rm{lat}}$ it can be used to define a renormalized coupling at scale $\mu = 1 / \sqrt{8t}$
\begin{equation}
  \label{eq:def_g2}
  \gGF(\mu = 1 / \sqrt{8t}) = \frac{1}{\cN} \vev{t^2 E(t)},
\end{equation}
where ``$a$'' is the lattice spacing, $t_{\rm{lat}}$ is dimensionless, and the energy density $E(t) = -\frac{1}{2}\mbox{ReTr}\left[G_{\mu\nu}(t) G^{\mu\nu}(t)\right]$ is calculated at flow time $t$ with an appropriate lattice operator.
We evolve the gradient flow with the Wilson plaquette term and use the usual ``clover'' or ``symmetric'' definition of $G_{\mu\nu}(t)$.
The normalization \cN is set such that $\gGF(\mu)$ agrees with the continuum \MSbar coupling at tree level.

If the flow time is fixed relative to the lattice size, $\sqrt{8t} = cL$ with $c$ constant, the scale of the corresponding coupling $\gc(L)$ is set by the lattice size.
Like the well-known Schr\"odinger functional (SF) coupling, $\gc(L)$ can be used to compute a step scaling function~\cite{Fodor:2012td, Fodor:2012qh, Fritzsch:2013je}.
The greater flexibility of the gradient flow running coupling is a significant advantage over the more traditional SF coupling.
A single measurement of the gradient flow will provide \gc for a range of $c$.
In our study we obtain \gc for all $0 \leq c \leq 0.5$ separated by $\de t_{\rm{lat}} = 0.01$.
Each choice of $c$ corresponds to a different renormalization scheme, which can be explored simultaneously on the same set of configurations~\cite{Fritzsch:2013je}.

The normalization factor \cN in finite volume has been calculated for anti-periodic boundary conditions (BCs) in refs.~\cite{Fodor:2012td, Fodor:2012qh}, and for SF BCs in \refcite{Fritzsch:2013je}.
In this work we use anti-periodic BCs, for which
\begin{align}
  \frac{1}{\cN} & = \frac{128\pi^2}{3(N^2 - 1)(1 + \de(c))} &
  \de(c) & = \vartheta^4\left(e^{-1 / c^2}\right) - 1 - \frac{c^4 \pi^2}{3},
\end{align}
where $\vartheta(x) = \sum_{n = -\infty}^{\infty} x^{n^2}$ is the Jacobi elliptic function.
For $0 \leq c \leq 0.3$ the finite-volume correction $\de(c)$ computed in \refcite{Fodor:2012td} is small, $|\de(c)| \leq 0.03$.
As explained in refs.~\cite{Fodor:2012td, Fodor:2012qh}, the RG \be function of \gGF is two-loop universal with SF BCs, but only one-loop universal with anti-periodic BCs.

At non-zero lattice spacing \gGF has cut-off corrections.
These corrections could be $\cO(a)$ for unimproved actions, and even $\cO(a)$-improved actions could have large $\cO(a^2 [\log a]^n)$-type corrections~\cite{Balog:2009yj, Balog:2009np}.
In existing numerical studies of staggered or $\cO(a)$-improved Wilson fermions the leading lattice corrections appear to be $\cO(a^2)$~\cite{Fritzsch:2013je, Sommer:2014mea},
\begin{equation}
  \label{eq:lat_g2}
  \gGF(\mu; a) = \gGF(\mu; a = 0) + a^2 \cC + \cO(a^4 [\log a]^n, a^4).
\end{equation}
It is possible to remove, or at least greatly reduce, the $\cO(a^2)$ corrections in \eq{eq:lat_g2} by defining
\begin{equation}
  \label{eq:t-shift}
  \gtGF(\mu; a) = \frac{1}{\cN} \vev{t^2 E(t + \tau_0 a^2)}\, ,
\end{equation}
where $\tau_0 \ll t / a^2$ is a small shift in the flow time.
In the continuum limit $\tau_0 a^2 \to 0$ and $\gtGF(\mu) = \gGF(\mu)$.

There are several possible interpretations of the $t$-shift in \eq{eq:t-shift}.
The gradient flow is an invertible smearing transformation, so one can consider $\tau_0$ as an initial flow that does not change the IR properties of the system but leads to a new action.
The gradient flow coupling \gtGF in \eq{eq:t-shift} is calculated for this new action.
Alternatively one can consider the replacement of $\vev{t^2 E(t)}$ with $\vev{t^2 E(t + \tau_0 a^2)}$ as an improved operator for the energy density.
In either case the $t$-shift changes the $\cO(a^2)$ term of $\gGF(\mu; a)$.
If we expand $\gtGF(\mu)$ in $\tau_0 a^2$,
\begin{equation}
  \label{eq:expand}
  \gtGF(\mu; a) = \frac{1}{\cN} \vev{t^2 E(t)} + \frac{a^2 \tau_0}{\cN} \vev{t^2 \frac{\partial E(t)}{\partial t}},
\end{equation}
and choose $\tau_0$ such that the second term in \eq{eq:expand} cancels the $a^2 \cC$ term in \eq{eq:lat_g2}, we remove the leading lattice artifacts
\begin{equation}
  \gtopt(\mu; a) = \gGF(\mu; a = 0) + \cO(a^4 [\log a]^n, a^4).
\end{equation}
Full $\cO(a^2)$ improvement through a systematic improvement program would require adding terms to improve the flow equation, the action, the boundary conditions, and the energy density operator $\vev{t^2 E(t)}$~\cite{Sommer:2014mea}.
Since our proposed improvement involves only a single parameter $\tau_0$, this $\tau_0$ itself must depend on other parameters, most importantly on $\gtGF(\mu)$ and on the bare coupling through the lattice spacing dependence of the term $\vev{t^2 \frac{\partial E(t)}{\partial t}}$ in \eq{eq:expand}.
Optimizing $\tau_0$ both in the renormalized and bare couplings could remove the predictive power of the method.
Fortunately, as we will see in the next section, our numerical tests indicate that it is sufficient to choose $\tau_0$ to be a constant or only weakly $\gtGF(\mu)$ dependent to remove most $\cO(a^2)$ lattice artifacts.

Since the gradient flow is evaluated through numerical integration, the replacement $\gGF \to \gtGF$ can be done by a simple shift of $t$ without incurring any additional computational cost.
The optimal $t$-shift \topt can be identified by a simple procedure when the gradient flow is used for scale setting, which we will consider in a future publication.
In this paper we concentrate on the step scaling function and find the \topt that removes the $\cO(a^2)$ terms of the discrete \be function corresponding to scale change $s$,
\begin{equation}
  \label{eq:beta_lat}
  \be_{\rm lat}(\gc; s; a) = \frac{\gtc(L; a) - \gtc(sL; a)}{\log(s^2)}.
\end{equation}

\section{Testing improvement with 4-flavor SU(3) gauge theory} 
\begin{figure}[btp]
  \centering
  \includegraphics[width=0.75\linewidth]{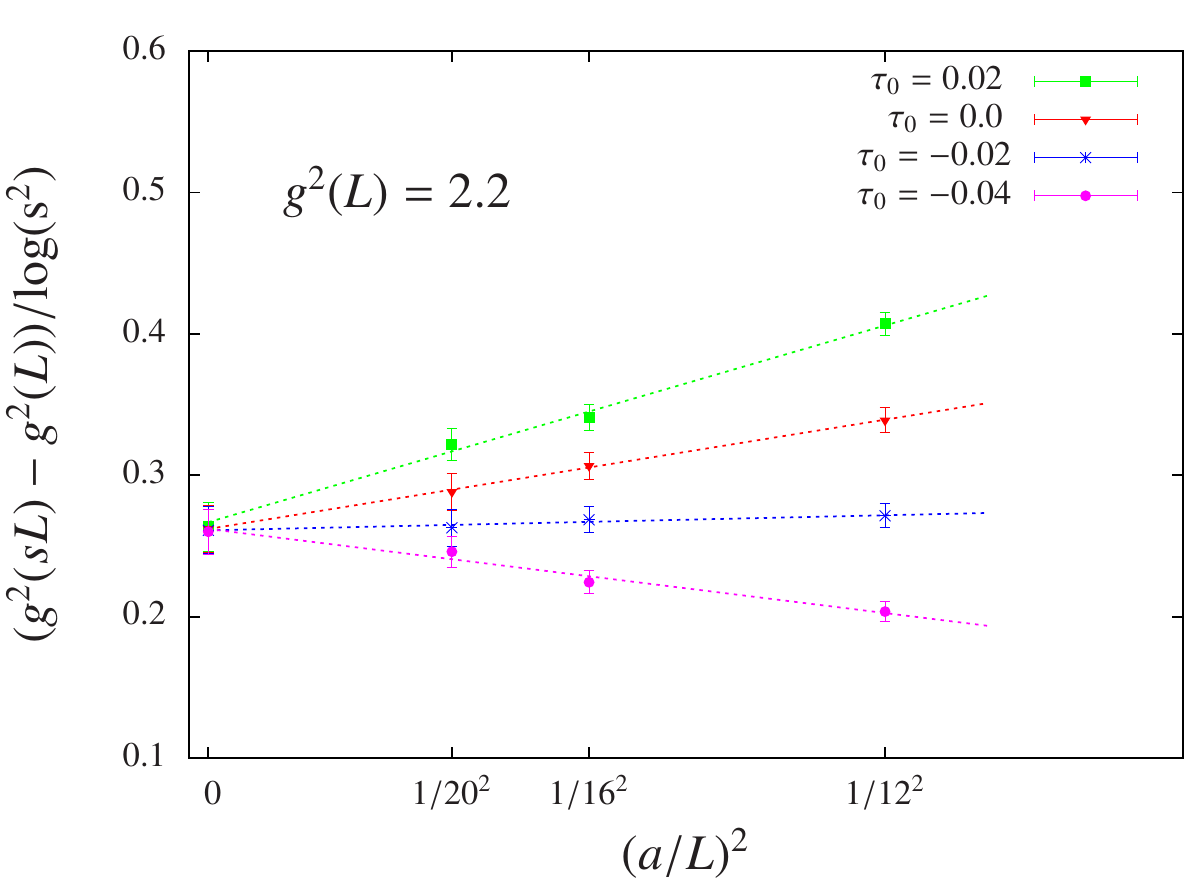}
  \caption{\label{fig:Nf4} Continuum extrapolations of the discrete $\be_{\rm lat}$ function of the $N_f = 4$ system at $\gtc(L) = 2.2$ with several different values of the $t$-shift coefficient $\tau_0$.  The dotted lines are independent linear fits at each $\tau_0$, which predict a consistent continuum value.}
\end{figure}
\begin{figure}[btp]
  \centering
  \includegraphics[width=0.75\linewidth]{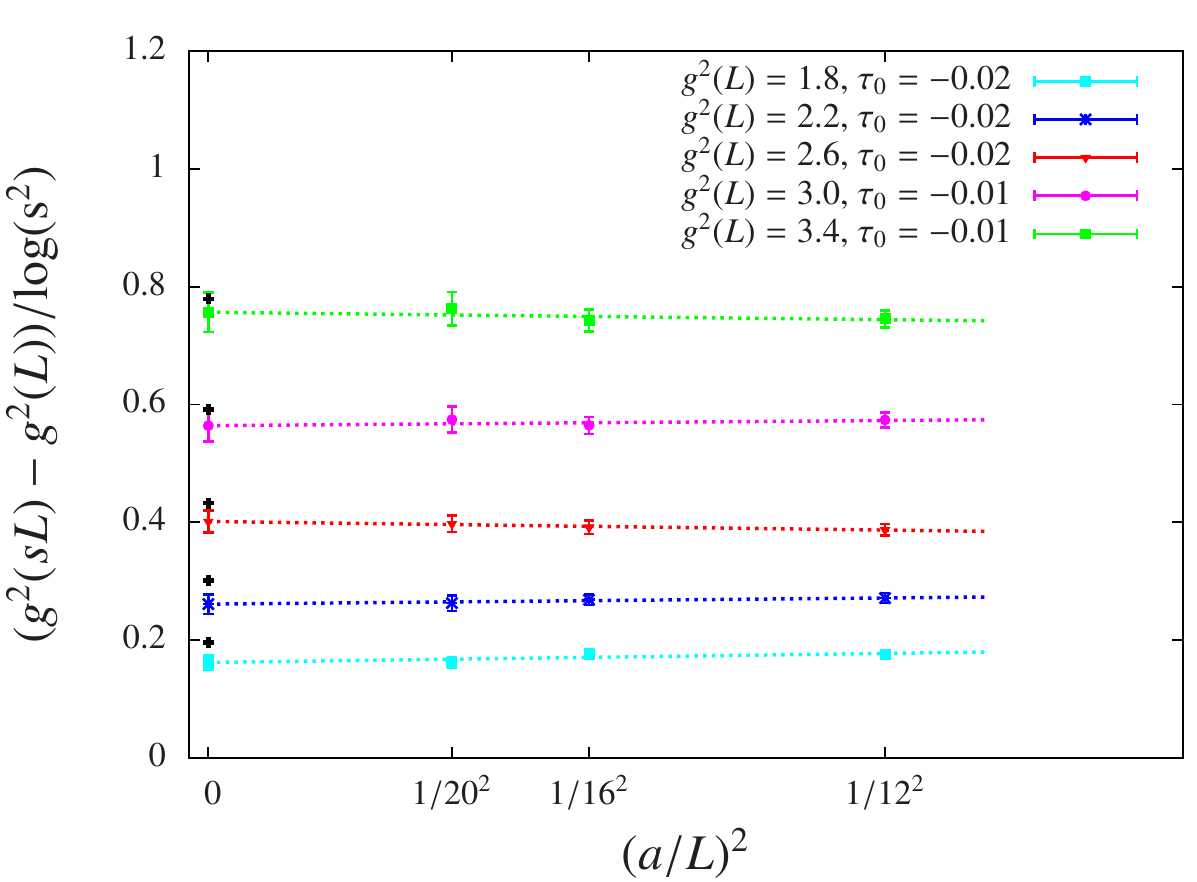}
  \caption{\label{fig:Nf4_many} Continuum extrapolations of the discrete $\be_{\rm lat}$ function of the $N_f = 4$ system for several different $\gtc(L)$ values.  For $\gtc(L) = 1.8$, 2.2 and 2.6 $\tau_0 = -0.02$ is near-optimal, while the larger couplings $\gtc(L) = 3.0$ and 3.4 require $\tau_0 = -0.01$ to remove most $\cO(a^2)$ effects.  The colored points at $(a / L)^2 = 0$ are the continuum extrapolated results, while the black crosses at $(a / L)^2 = 0$ show the corresponding two-loop perturbative predictions.}
\end{figure}

We illustrate the $t$-shift improvement with the $N_f = 4$ SU(3) system.
This theory was recently studied by refs.~\cite{Fodor:2012td, Fodor:2012qh} using gradient flow step scaling with staggered fermions.
The 4-flavor SF running coupling was previously considered in \refcite{Tekin:2010mm} using $\cO(a)$-improved Wilson fermions, and in \refcite{PerezRubio:2010ke} using staggered fermions.
In our calculations we use nHYP-smeared~\cite{Hasenfratz:2001hp, Hasenfratz:2007rf} staggered fermions and a gauge action that includes an adjoint plaquette term in order to move farther away from a well-known spurious fixed point in the adjoint--fundamental plaquette plane~\cite{Cheng:2011ic}.
As in \refcite{Fodor:2012td} we impose anti-periodic BCs in all four directions, which allows us to carry out computations with exactly vanishing fermion mass, $m = 0$.
For the discrete \be function we consider the scale change $s = 3 / 2$ and compare lattice volumes $12^4 \to 18^4$, $16^4 \to 24^4$ and $20^4 \to 30^4$.
We accumulated 500--600 measurements of the gradient flow coupling, with each measurement separated by 10 molecular dynamics time units (MDTU), at 7--8 values of the bare gauge coupling on each volume.
We consider the $c = 0.25$ scheme, as opposed to $c = 0.3$ used in \refcite{Fodor:2012td}, because smaller $c$ gives better statistics at the expense of larger lattice artifacts.
As discussed above, we aim to reduce these lattice artifacts through the non-perturbative improvement we have introduced.
We follow the fitting procedure described in \refcite{Tekin:2010mm}.

Full details of this study will be presented in \refcite{Cheng:2014}.
Here we provide a representative illustration of the $t$-shift optimization.
Figure~\ref{fig:Nf4} shows the dependence of the discrete \be function on $(a / L)^2$ when $\gtc(L) = 2.2$ with several values of the $t$-shift parameter $\tau_0$.
The red triangles correspond to no improvement, $\tau_0 = 0$.
The data are consistent with linear dependence on $a^2$ and extrapolate to 0.262(17), about $2\sigma$ below the two-loop perturbative value of 0.301.
The slope of the extrapolation is already rather small, $b = 11(3)$.
By adding a small shift this slope can be increased or decreased.
With $\tau_0 = -0.02$ no $\cO(a^2)$ effects can be observed -- the corresponding slope is $b = 1.5(3.1)$ -- and we identify this value as near the optimal $\topt$.
The data at different $\tau_0$ extrapolate to the same continuum value, even when the slope $b$ is larger than that for $\tau_0 = 0$.
This is consistent with the expectation that the $t$-shift changes the $\cO(a^2)$ behavior of the system but does not affect the continuum limit.
Since our action produces relatively small $\cO(a^2)$ corrections even without improvement, the $t$-shift optimization has little effect on the continuum extrapolation, though the consistency between different values of $\tau_0$ is reassuring.

It is interesting that the cut-off effects in our unimproved results, characterized by the slope $b$ of the red triangles in \fig{fig:Nf4}, are more than three times smaller than those shown in fig.~4 of \refcite{Fodor:2012td}.
This difference grows to about a factor of four when we consider the larger $c = 0.3$ used in that study, suggesting that the $t$-shift optimization could have a more pronounced effect with the action used in \refcite{Fodor:2012td}.
The cause of the reduced lattice artifacts with our action is not obvious.
Both our action and that used by \refcite{Fodor:2012td} are based on smeared staggered fermions, though we use different smearing schemes.
The different smearing might have an effect, as might the inclusion of the adjoint plaquette term in our gauge action.
This question is worth investigating in the future.

In principle \topt could be different at different \gc couplings but in practice we found little variation.
Figure~\ref{fig:Nf4_many} shows near-optimal continuum extrapolations of the discrete \be function at several values of $\gtc(L)$.
At each $\gtc(L)$ the continuum extrapolated result is consistent within $\sim$$2\sigma$ with the two-loop perturbative prediction, denoted by a black cross in \fig{fig:Nf4_many}.
Comparable consistency with perturbation theory was found in previous studies~\cite{Tekin:2010mm, PerezRubio:2010ke, Fodor:2012td, Fodor:2012qh}.

\section{Infrared fixed point in 12-flavor SU(3) gauge theory} 
\begin{figure}[btp]
  \centering
  \includegraphics[width=0.75\linewidth]{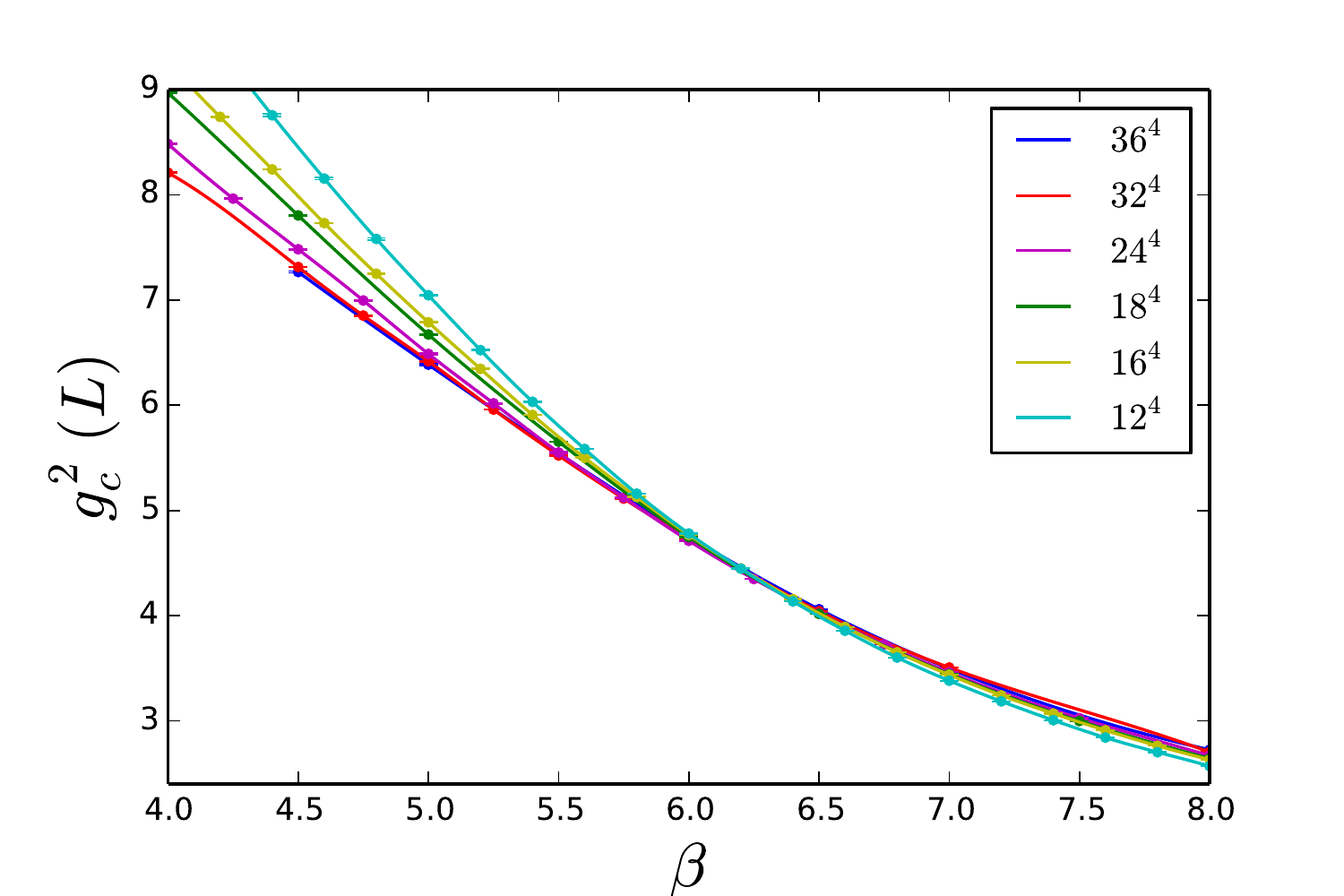}
  \caption{\label{fig:gradient_flow}The $N_f = 12$ running coupling $\gc(L)$ versus the bare coupling $\be_F$ on several volumes, for $c = 0.2$.  Crossings between results from different volumes predict the finite volume IRFP coupling $\gstar(L)$ in this scheme.}
\end{figure}

We use the same lattice action with $N_f = 12$ as with $N_f = 4$ and consider six different volumes: $12^4$, $16^4$, $18^4$, $24^4$, $32^4$ and $36^4$.
This range of volumes allows us to carry out step scaling analyses with scale changes $s = 4 / 3$, $3 / 2$ and 2.
As for $N_f = 4$ we performed simulations in the $m = 0$ chiral limit with anti-periodic BCs in all four directions.
Depending on the volume and bare coupling $\be_F$ we accumulated 300--1000 measurements of the gradient flow coupling \gc for $0 \leq c \leq 0.5$, with 10 MDTU separating subsequent measurements.
Here we will consider only $c = 0.2$.
Full details of our ensembles and measurements, studies of their auto-correlations, and additional analyses for $c = 0.25$ and 0.3 will appear in \refcite{Cheng:2014}.
The choice of $c = 0.2$ minimizes the statistical errors, and we find the IRFP in this scheme to be at a weaker coupling than for larger $c$, which is numerically easier to reach.
The typical trade-off for these smaller statistical errors would be larger cut-off effects, but as discussed in previous sections these cut-off effects can be reduced by our non-perturbative improvement.

\begin{figure}[btp]
  \centering
  \includegraphics[width=0.75\linewidth]{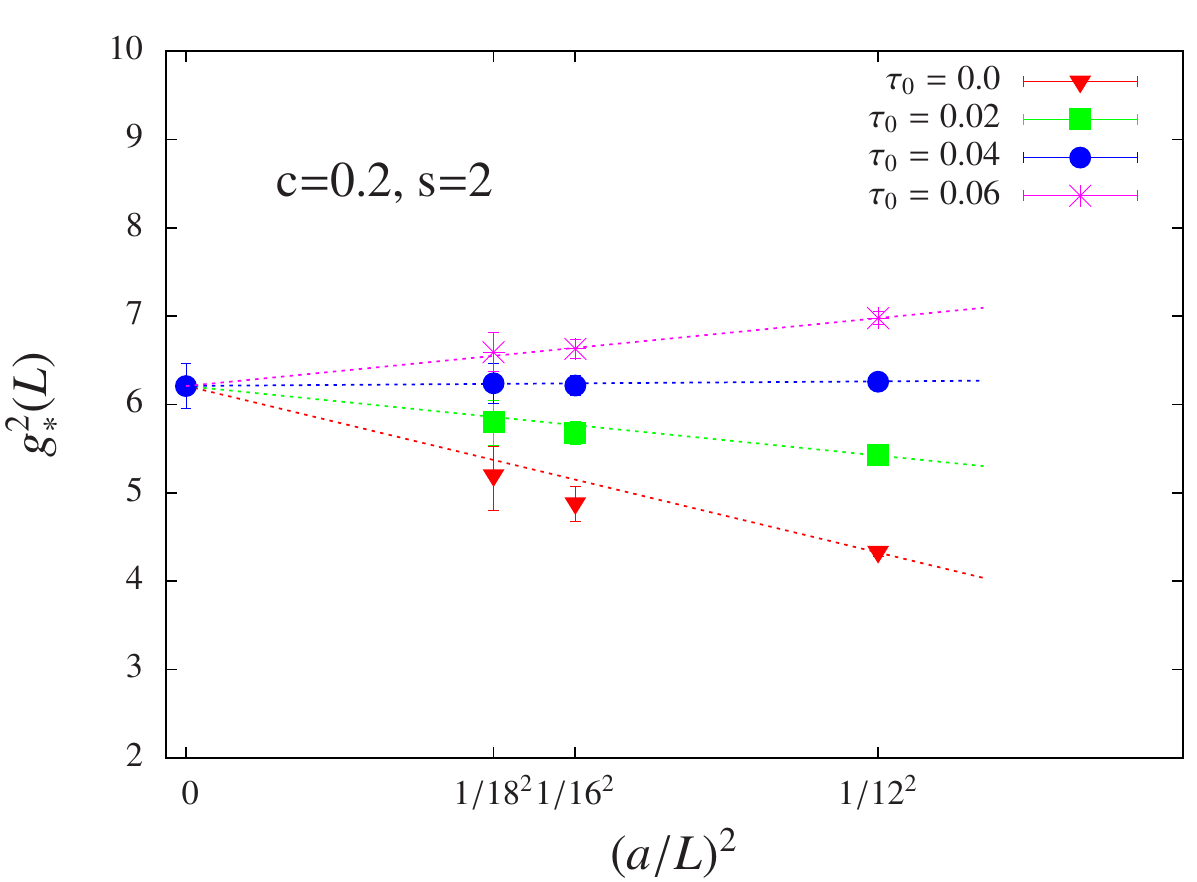}
  \caption{\label{fig:Nf12}Continuum extrapolations of the 12-flavor finite volume IRFP $\gstar(L)$, with several different $t$-shift coefficients $\tau_0$ for fixed scale change $s = 2$.  The dotted lines are a joint linear fit constrained to have the same $(a / L)^2 = 0$ intercept, which gives $\gstar = 6.21(25)$.}
\end{figure}

Figure~\ref{fig:gradient_flow} shows the running coupling $\gc(L)$ as the function of the bare gauge coupling $\be_F$ for different volumes.
The interpolating curves are from fits similar to those in \refcite{Tekin:2010mm}.
The curves from different volumes cross in the range $6.0 \leq \be_F \leq 6.5$.
The crossing from lattices with linear size $L$ and $sL$ defines the finite-volume IRFP coupling $\gstar(L; s)$:
\begin{equation}
  \gc(L) = \gc(sL) \implies \gstar(L; s) = \gc(L).
\end{equation}
If the IRFP exists in the continuum limit then the extrapolation
\begin{equation}
  \lim_{(a / L)^2 \to 0} \gstar(L; s) \equiv \gstar
\end{equation}
has to be finite and independent of $s$.\footnote{We thank D.~N\'ogr\'adi for useful discussions of the continuum limit.}
Figure~\ref{fig:Nf12} illustrates the continuum extrapolation of $\gstar(L)$ with scale change $s = 2$ for various choices of the $t$-shift parameter $\tau_0$.
The red triangles correspond to no shift, $\tau_0 = 0$.
Their $(a / L)^2 \to 0$ continuum extrapolation has a negative slope, and the leading lattice cut-off effects are removed with a positive $t$-shift, $\topt \approx 0.04$.
A joint linear extrapolation of the $\tau_0 = 0$, 0.02, 0.04 and 0.06 results, constrained to have the same continuum limit at $(a / L)^2 = 0$, predicts $\gstar = 6.21(25)$.
However, these results all come from the same measurements, and are therefore quite correlated.
While it is an important consistency check that the continuum limit does not change with $\topt$, just as for $N_f = 4$, the uncertainty in the continuum-extrapolated \gstar from this joint fit is not reliable.

\begin{figure}[btp]
  \centering
  \includegraphics[width=0.75\linewidth]{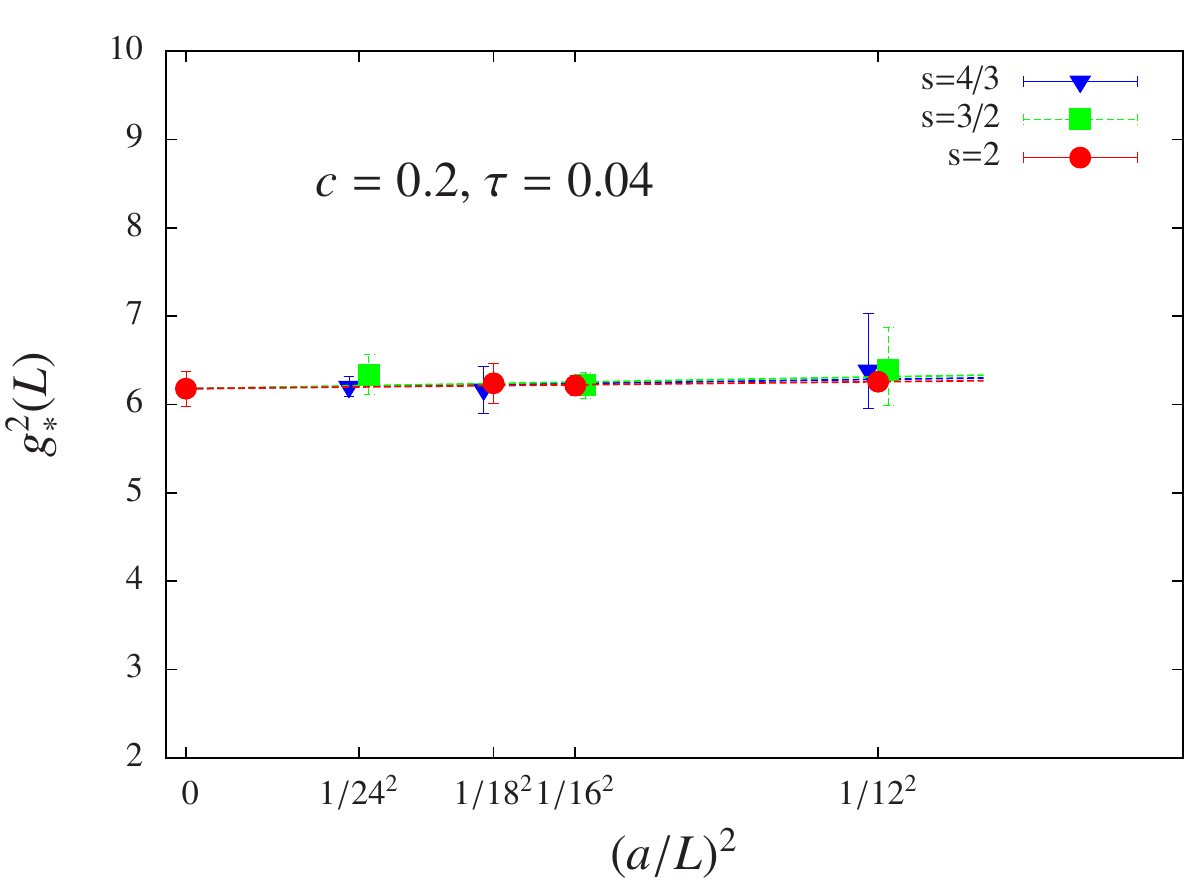}
  \caption{\label{fig:Nf12_all}Continuum extrapolations of the 12-flavor finite volume IRFP $\gstar(L)$, with several different scale changes for the near-optimal improvement coefficient $\topt \approx 0.04$.  The $s = 4 / 3$ and $3 / 2$ data points are horizontally displaced for greater clarity.  The dashed lines are a joint linear fit constrained to have the same $(a / L)^2 = 0$ intercept, which gives $\gstar = 6.18(20)$.}
\end{figure}

Instead, we should consider only the results with the near-optimal $\topt \approx 0.04$.
As we show in \fig{fig:Nf12_all}, $\topt \approx 0.04$ is also near-optimal for scale changes $s = 3 / 2$ and $4 / 3$.
None of these results have any observable $\cO(a^2)$ effect, making the extrapolation to the continuum very stable.
Each scale change predicts a continuum IRFP for $N_f = 12$.
The three sets of results in \fig{fig:Nf12_all} come from matching different volumes, making a joint fit legitimate.
This continuum extrapolation predicts that the IR fixed point is located at renormalized coupling $\gstar = 6.18(20)$ in the $c = 0.2$ scheme.


\section{Conclusion and summary} 
We have considered step scaling based on the gradient flow renormalized coupling, introducing a non-perturbative $\cO(a^2)$ improvement that removes, or at least greatly reduces, leading-order cut-off effects.
This phenomenological improvement increases our control over the extrapolation to the infinite-volume continuum limit, as we demonstrated first for the case of SU(3) gauge theory with $N_f = 4$ massless staggered fermions.
Turning to $N_f = 12$, we found that the continuum limit was well defined and predicted an infrared fixed point even without improvement.
Applying our proposed improvement reinforced this conclusion by removing all observable $\cO(a^2)$ effects.
For the finite-volume gradient flow renormalization scheme defined by $c = 0.2$, we find the continuum conformal fixed point to be located at $\gstar = 6.18(20)$.

The 12-flavor system has been under investigation for some time, and other groups have studied its step scaling function~\cite{Lin:2012iw, Appelquist:2007hu, Appelquist:2009ty, Itou:2012qn}.
However, this work is the first to observe an IRFP in the infinite-volume continuum limit.
There are likely several factors contributing to this progress.
While we did not invest more computer time than other groups, we have employed a well-designed lattice action.
The adjoint plaquette term in our gauge action moves us farther away from a well-known spurious fixed point, while nHYP smearing allows us to simulate at relatively strong couplings.
The gradient flow coupling itself appears to be a significant improvement over other schemes,\footnote{C.-J.~D.~Lin has told us about dramatic improvements in auto-correlations when using the gradient flow coupling compared to the twisted Polyakov loop coupling of \refcite{Lin:2012iw}.} and our non-perturbative improvement also contributes to obtaining more reliable continuum extrapolations.

Our non-perturbative improvement is general and easy to use in other systems.
It does not rely on the lattice action or fermion discretization, though we suspect that the improvement may not be effective if there are $\cO(a)$ artifacts, e.g.\ for unimproved Wilson fermions.
Since $\cO(a)$-improved lattice actions are standard, this does not appear to be a practical limitation.
We look forward to seeing our proposal applied both to QCD and to other conformal or near-conformal systems.

\section*{Acknowledgments} 
We thank J.~Kuti for suggesting a perturbative interpretation of our proposed improvement, and D.~N\'ogr\'adi for useful discussions of the continuum limit.
Both J.~Kuti and C.-J.~D.~Lin have kindly told us about their groups' ongoing investigations of the $N_f = 12$ gradient flow step scaling function.
A.~H.\ is grateful for the hospitality of the Brookhaven National Laboratory HET group and of the Kobayashi--Maskawa Institute at Nagoya University during her extended visits, as well as the Japan Society for the Promotion of Science Fellowship that made the latter visit possible.
This research was partially supported by the U.S.~Department of Energy (DOE) through Grant Nos.~DE-SC0010005 (A.C., A.H., Y.L.\ and D.S.), DE-SC0008669 and DE-SC0009998 (D.S.), and by the DOE Office of Science Graduate Fellowship Program under Contract No.~DE-AC05-06OR23100 (G.P.).
Our code is based in part on the MILC Collaboration's public lattice gauge theory software.\footnote{\texttt{http://www.physics.utah.edu/$\sim$detar/milc/}}
Numerical calculations were carried out on the University of Colorado HEP-TH cluster and on the Janus cluster partially funded by U.S.~National Science Foundation (NSF) Grant No.~CNS-0821794; at Fermilab under the auspices of USQCD supported by the DOE; and at the San Diego Computing Center through XSEDE supported by NSF Grant No.~OCI-1053575.

\bibliographystyle{utphys}
\bibliography{gradflow}
\end{document}